\documentclass[prb,twocolumn,notitlepage,showpacs,amsmath,amstex,amssymb,citeautoscript,longbibliography,groupedaddress,]{revtex4-1}
\pdfoutput=1

\usepackage{natbib}
\usepackage[english]{babel}
\usepackage{letltxmacro}
\usepackage{latexsym}
\LetLtxMacro{\ORIGselectlanguage}{\selectlanguage}
\makeatletter
\DeclareRobustCommand{\selectlanguage}[1]{%
  \@ifundefined{alias@\string#1}
    {\ORIGselectlanguage{#1}}
    {\begingroup\edef\x{\endgroup
       \noexpand\ORIGselectlanguage{\@nameuse{alias@#1}}}\x}%
}
\newcommand{\definelanguagealias}[2]{%
  \@namedef{alias@#1}{#2}%
}
\makeatother
\definelanguagealias{en}{english}
\definelanguagealias{English}{english}
\usepackage{graphicx}
\usepackage{amsmath}
\usepackage{mathtools}
\usepackage{amsfonts}
\usepackage{amssymb,bbding}
\usepackage{bm}
\usepackage{color}
\usepackage[percent]{overpic}
\usepackage{soul} 
\usepackage{amssymb}
\usepackage{wasysym}
\usepackage{dsfont}
\usepackage{float}
\usepackage{braket}
\usepackage{subfiles}
\usepackage{epstopdf}

\usepackage[normalem]{ulem}

\usepackage{tikz}
\usetikzlibrary{positioning}

\usepackage{blkarray}
\usepackage{multirow}

\usepackage{hyperref}
\hypersetup{
    bookmarks=false,         
    unicode=false,          
    pdftoolbar=false,        
    pdfmenubar=true,        
    pdffitwindow=false,     
    pdfstartview={FitH},    
    pdftitle={Breakdown of energy flux quantization in a dissipating quantum Hall edge},    
    pdfauthor={A. Goremykina, A.Borin, E.Sukhorukov},     
    pdfsubject={},   
    pdfcreator={},   
    pdfproducer={}, 
    pdfkeywords={quantum Hall} {energy flux} {dissipation}{neutral modes}, 
    pdfnewwindow=true,      
    colorlinks=true,       
    linkcolor=black,          
    citecolor=blue,        
    filecolor=magenta,      
    urlcolor=blue           
}

\newcommand{\be}{\begin{equation}}
\newcommand{\ee}{\end{equation}}
\newcommand{\bea}{\begin{eqnarray}}
\newcommand{\eea}{\end{eqnarray}}

\newcommand{\vp}{\varphi}
\newcommand{\ve}{\varepsilon}
\newcommand{\mc}{\mathcal}
\newcommand{\dvp}{\delta\varphi}
\newcommand{\sx}{\sigma_{xx}}
\newcommand{\sy}{\sigma_{xy}}

\begin{document}
\title{Heat current in a dissipative quantum Hall edge}

\author{Anna Goremykina,  Artem Borin,  Eugene Sukhorukov}
\affiliation{
D\'epartement de Physique Th\'eorique, Universit\'e de Gen\`eve, CH-1211 Gen\`eve 4, Switzerland}

\date{\today}
\begin{abstract}
We explore the heat current in the quantum Hall edge at filling factors $\nu = 1$ and $\nu = 2$ in the presence of dissipation. Dissipation arises in the compressible strip forming at the edge in presence of a smooth confining potential. Such strip was predicted to host an infinite number of hydrodynamic neutral modes, which however were never observed. A possible explanation may be in their dissipative nature, which was not fully considered before. Heat transport measurements are capable of detecting neutral modes and experiment [H. le Sueur et al., Phys. Rev. Lett. 105, 056803 (2010)] at $\nu = 2$ captured additional degrees of freedom transferring heat at the edge. Surprisingly, the breakdown of heat current quantization has been found. We conjecture that the aforementioned dissipative modes might be responsible for this behavior. We build a low-energy effective model and show that the lowest hydrodynamic mode carries a portion of the heat flux quantum which is the same both at  $\nu  = 1$ and $\nu = 2$.
Although our results are consistent with the experiment, a microscopic model of dissipation is needed to confirm the prediction of the low-energy approximation.
\end{abstract}
\maketitle

\section{Introduction}
The heat transport experiments are able to provide insights into physics when conventional charge conductance measurements are powerless~\cite{banerjee2017, KaneFisher97, Jezouin601, PhysRevLett.97.056601, PhysRevLett.68.3765, schwab2000, PhysRevB.93.245427, gurman2012}. They have been particularly useful in understanding a structure and intriguing properties of quantum Hall (QH) edge states both at integer and fractional bulk fillings \cite{heat_transport_nu1, Venkatachalam, banerjee2017, banerjee2018, PhysRevLett.121.026801, pierre}. Indeed, due to effects of interactions, disorder, smoothness of the confining potential and the finite temperatures the density profile of the QH edge was predicted to acquire a non-trivial form \cite{PhysRevLett.64.220, PhysRevLett.67.2060, PhysRevB.43.11025}, resulting in the additional neutral counter-propagating excitations \cite{PhysRevLett.72.2624, PhysRevLett.111.246803, PhysRevLett.91.036802, PhysRevB.68.035332, PhysRevB.90.165104, PROTOPOPOV2017287, PhysRevB.92.245437, PhysRevLett.88.056802, PhysRevB.52.16357}. Later on, the measurements of a thermal conductance allowed for the detection of these upstream modes \cite{bid2010, PhysRevLett.113.266803, gurman2012, PhysRevLett.108.226801, PhysRevLett.107.036805, banerjee2017}, opening a way to testing the existing theories of the edge. This is based on the quantization of the thermal current independently of the nature of the modes that transmit it~\cite{KaneFisher97}.  A lot of attention is drawn to the fractional QH regimes described by non-Abelian symmetries, which give rise to exotic states \cite{banerjee2018,MOORE1991362, PhysRevLett.121.026801, PhysRevLett.59.1776}. However, the interference experiments that might provide the evidence of their anyonic statistics are believed to be hindered by such neutral modes~\cite{PhysRevLett.117.276804}, the idea recently supported by a systematic experimental study \cite{PhysRevLett.122.246801}. 

The integer QH edge was also predicted by ~\textcite{AG} to host neutral (AG) excitations. Their approach is based on the electrostatic theory of the edge~\cite{PhysRevB.46.4026, PhysRevB.49.8227}, which states that in the presence of screening and smooth confining potential the charge density profile consists of alternating compressible and incompressible regions, an experimentally confirmed picture  \cite{PhysRevX.4.011014}.  Within an exactly solvable hydrodynamic model of a compressible strip profile it was demonstrated that such strip hosts an infinite number of downstream neutral excitations. So far a direct observation of these modes has remained elusive. What has been observed is a significant leakage of the injected energy from the edge channels at integer $\nu \in [1,3]$~\cite{heat_transport_nu1, Venkatachalam}, suggesting the presence of additional degrees of freedom to which the energy might be redistributed. 

\begin{figure}[h!]
\center
\includegraphics[scale = 0.7]{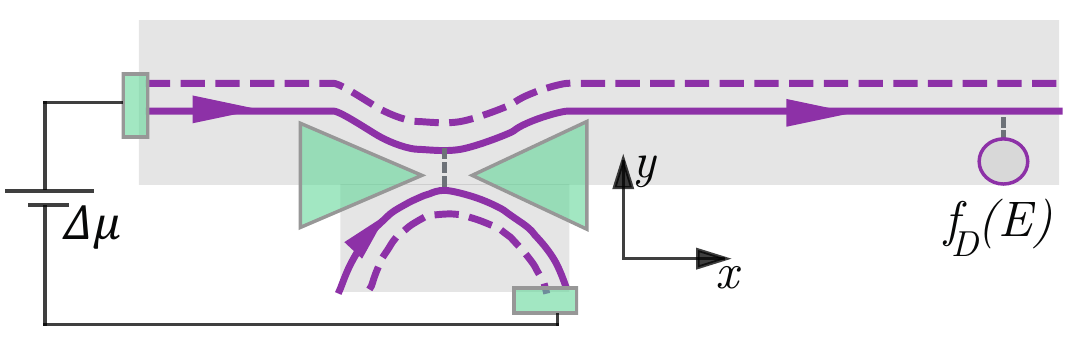}
\caption{A scheme of the experimental set-up [\onlinecite{pierre}]. An outer edge channel at $\nu = 2$ is driven out-of-equilibrium by the quantum point contact (triangles) and the energy distribution of the created excitation is extracted from probing the tunneling density of states $\propto \partial f_D(\varepsilon)/\partial\ve$ by the quantum dot placed downstream. \label{fig:scheme}}\end{figure}

A systematic study of the heat transport at $\nu = 2$ was made in Ref.~[\onlinecite{pierre}]. In this system, Fig.~\ref{fig:scheme}, electrons are injected into the outer edge channel via a quantum point contact and the energy distribution in this channel is subsequently probed downstream after different lengths of propagation. 
The first observation is that the energy does not leak into the bulk, since one finds the saturation of the local temperature even after large propagation lengths of $10\mu m$. However, the measured temperature is lower than is predicted by the equipartition of the initial energy between the two modes.
The temperature is extracted from the calculation of the energy flux $J_E = \int\frac{ d\varepsilon}{2\pi} \varepsilon \Big[f(\varepsilon) - \theta(\mu - \varepsilon)\Big]$, where $\mu$ is the electrochemical potential and $f(\varepsilon)$ is a measured energy distribution.  Its value is expected to acquire the energy flux quantum 
\begin{equation}
J_q = \frac{\pi T^2}{12}.
\end{equation}
Instead, it misses around $13\%$ of this number. This outcome is also inconsistent with the three modes structure of the edge, since then the injected energy would have been equipartitioned between these three modes.
But what if the additional mode carried less than $J_q$, say due to the presence of dissipation? A compressible strip, that is predicted to host neutral modes, allows for a transverse current  proportional to the diagonal conductivity $\sigma_{xx}$, which results in a dissipative term in the energy spectrum of these excitations (which we show below). Previously, effects of dissipation did not receive a full consideration in the model~[\onlinecite{AG}]. However, dissipation was observed\cite{bocquillon} in the spectrum of the neutral mode at $\nu = 2$  and was also predicted~\cite{PhysRevB.97.115418} to account for the experimental finding of Ref.~[\onlinecite{main_roche}].

In this paper we concentrate on the low-energy limit and demonstrate how the introduced dissipation in the compressible strip alters the spectrum of edge excitations at $\nu = 1$ and generalize our approach for $\nu = 2$. Under the assumption that no diffusion is present and that $\sigma_{xx}$ dominates over  $\sigma_{xy}$ in the strip we find that the additional AG mode carries a portion of flux quantum proportional to $\sigma_{xy}/\sigma_{xx}$. We show that in the presence of only one such strip the result is universal also at $\nu = 2$, which is a consequence of the low-energy character of our model. We note however that the existence of the high energy cut-off, which is a requirement of our model, is essential for the AG mode to have a non-quantized heat current. While our predictions are consistent with the experimental observations of Refs.~[\onlinecite{pierre}, \onlinecite{bocquillon}], we remark that in order to understand whether a low-energy approach is fully valid in such a problem, a microscopic model of dissipation has to be constructed. We reserve this direction for the future work.

The structure of the paper is the following. In Sec.~\ref{sec:model} we provide a model of a dissipative compressible strip and find the spectrum of the edge excitations at $\nu = 1$. Then, in Sec.~\ref{sec:flux} we explain how the energy flux can be calculated within the aforementioned model. The answer is represented as a sum of various density-density correlation functions, which can be found from the Fluctuation-dissipation theorem (FDT), Sec.~\ref{sec:FDT}. In the latter section, we also calculate the expression for the heat current of the two modes at $\nu = 1$ and explain the logics behind the result for the heat current of the AG mode. We generalize the above formalism for $\nu = 2$ in Sec.~\ref{sec:2} and provide final remarks in the discussion section ~\ref{sec:disc}.

\section{Compressible strip model \label{sec:model}}

\begin{figure}[h!]
\center
\includegraphics[scale = 0.5]{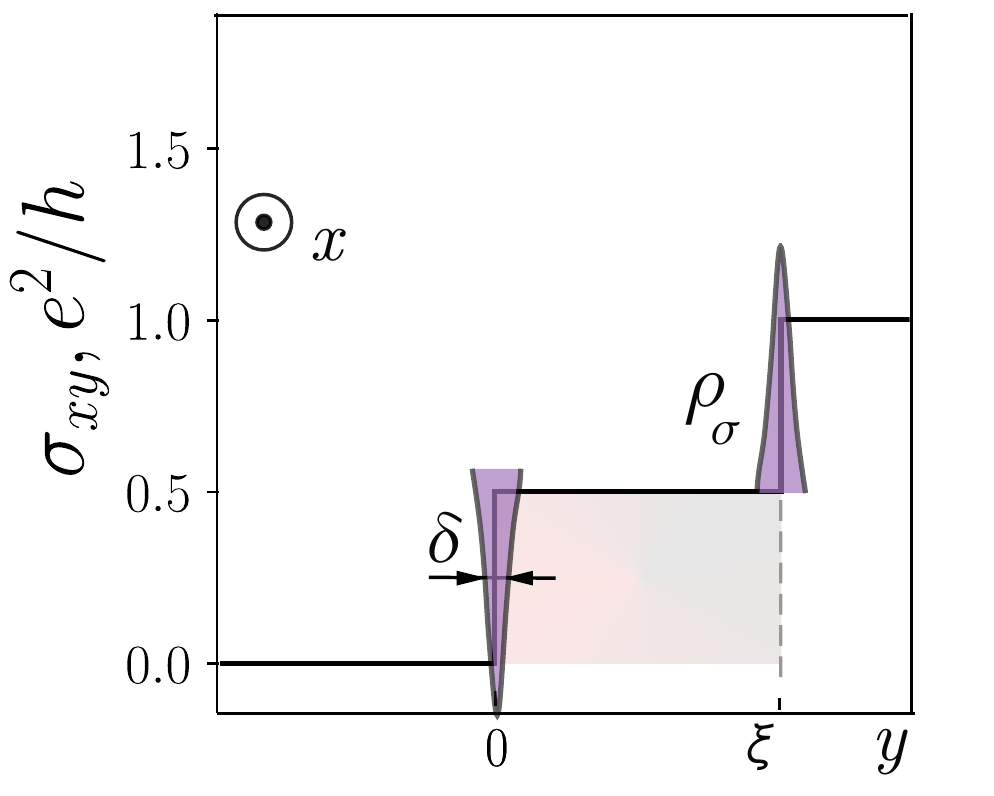}
\caption{A simplified Hall conductivity profile of the edge cross-section at $\nu = 1$. In presence of screening and smooth confining potential the QH plateaus admit compressible regions, which we display only for the first plateau as a $1/2$ step of the width $\xi$ (shaded region). We have also schematically depicted the neutral mode with the amplitude $\rho_{\sigma}$, that it hosted by the dissipative strip. The effective width of the charge density localized in the edge channels is denoted by $\delta$. \label{fig:density}}
\end{figure}

The density profile of a compressible strip forming close to the boundary of the edge is described by a smooth function which ends up with a plateau, corresponding to the density $1/2\pi \ell^2_B$, where $\ell_B$ is a magnetic length. Instead of taking care of a possibly infinite number of AG modes residing in it, we note that the hydrodynamic picture of the edge breaks down at short wave lengths, so that ultimately only a few such modes can exist. To capture the lowest neutral mode of the compressible region we introduce a half jump \footnote{Experiments \citep{klitzing, Suddards_2012, doi:10.1098/rsta.2011.0198} show that the peak of $\sigma_{xx}$ is situated approximately in the middle between the two $\sigma_{xy}$ plateaus. We thus model the compressible strip which hosts dissipation as a symmetric jump in the profile of $\sigma_{xy}$. Any small deviation results  in the perturbative correction since the leading order modes stay the same.} in the profile of $\sigma_{xy}(y)$ as presented in Fig.~\ref{fig:density} for $\nu = 1$. Any jump in $\sigma_{xy}(y)$ corresponds to the accumulation of the charge density. Hence,  in- and out-of-phase oscillations of the charges with amplitudes $\rho_{1,2}$ at the respective boundaries of the strip at $y=0$ and $y = \xi$ correspond to the charged and neutral excitations 
\begin{equation}
\rho_{c,\sigma}(k, \omega, y) = \rho_2(k, \omega)\delta(y-\xi) \pm \rho_1(k, \omega) \delta(y),
\end{equation}
 whose symmetric form is due to the symmetric profile of $\sigma_{xy}(y)$.
In this case, the charged mode is a usual magnetoplasmon, which is insensitive to the dissipation, while the neutral mode acquires a dominant dissipative contribution to its spectrum. To show this, let us write down a continuity equation for the total charge $\rho(k, y)$ in the compressible region: 
\begin{align}\label{1st_eq}
\omega \rho(k, y) = i \Big[&-i k (\partial_y \sigma_{xy}(y)) +
(\partial_y \sigma_{xx}(y))\partial_y \nonumber\\
& +\sigma_{xx}(y)(\partial^2_y - k^2) 
\Big]\varphi(k, y),
\end{align}
with the potential $\vp(k,y)$ created by the charges
\begin{equation}\label{ch5_phi}
\vp(k, y) = 2\pi \Big[ U(|k y|) \rho_1(k) + U(|k (y-\xi)|)\rho_2(k)\Big] 
\end{equation}
and $U(|k y|)$ describing a Coulomb interaction. We now make a few assumptions. We are primarily interested in the low-energy physics and thus neglect the $k^2$ term in Eq.~\eqref{1st_eq}. Basically, it means that we ignore the diffusion along the edge due to $\sigma_{xx}$. Experimentally, this is a natural assumption since the observed transport is chiral. It also seams reasonable if we imagine that the strip is non-uniform along the $x$ direction in the long-wave length limit and thus there is no contribution to the longitudinal current from $\sigma_{xx}$ term. Next, we assume a constant electric field inside the strip, i.e.  $ U'(|ky|)|_{y=\delta}\! =\! U'(|k y|)|_{y = \xi}$, where $\delta$ is an effective width of the channel. This approximation follows from the low-energy limit and can be understood from the Eq.~\eqref{1st_eq}, when the $\omega$ term is neglected.
Substituting
\begin{align}
&\sigma_{xx}(y) = \sigma_{xx}\theta(y)\theta(\xi-y), \\ 
&\sigma_{xy} (y) = \frac{\sigma_{xy}}{2}\Big[\theta(y) + \theta(y- \xi)\Big]
\end{align}
 into Eq.~\eqref{1st_eq} we get a system
\begin{subequations}\label{system}
\begin{eqnarray}
\omega \rho_1 = k\sigma_{xy}\vp(k,\delta)/2 + i\sigma_{xx} \partial_y\vp(k,y)|_{\delta},\\
\omega \rho_2 = k\sigma_{xy}\vp(k,\xi)/2 - i\sigma_{xx} \partial_y\vp(k,y)|_{\xi}.
\end{eqnarray}
\end{subequations}
which is easily diagonalized by the modes $\rho_{c,\sigma}(k)$ with the spectrum
\begin{align}
\omega_c(k) = k v_c, \quad \omega_{\sigma}(k) = k v_{\sigma} - 2i\frac{\sigma_{xx}}{\sigma_{xy}} \ve_{0}.
\end{align}
We assume that the Coulomb interaction is screened at distance $D$,  which is much larger than the width of the strip and much smaller than  the wavelength. Therefore, the velocities
\begin{equation}
v_{c,\sigma} \!= \! 2\pi \sigma_{xy}\Big[U(\delta/D) \pm U(\xi/D)\Big]/2,
\end{equation}
do not have dispersion. Such screening is naturally present in the experiment [\onlinecite{pierre}] due to various metallic gates.
We also introduced the energy parameter $\ve_0 \equiv-2\pi \sigma_{xy} U'(y/D)|_{y = \delta}$ which can be written in the low energy limit $k \ll 1/\xi$ as  
\begin{equation}
\quad \ve_0 = 2 v_{\sigma}/\xi.
\end{equation}
Energy $\ve_0$ is thus related to the ultra-violet cut-off $1/\xi$ for the wave vector $k$, and the dissipative term contains a parameter $\sigma_{xx}/\sigma_{xy}$ defining its dominant contribution in the compressible strip.

\section{Energy flux: general approach\label{sec:flux}}

So far we have used classical equations. To incorporate quantum effects into the definition of the energy flux $J_E$ carried by the edge at the thermal equilibrium, we write it as a symmetric product of the operators for the potential energy and the current along the edge $j_x(y)$:
\begin{align}\label{flux_def}
J_E\! =\! \! \int\! \frac{d\omega dk dk'}{(2\pi)^3} \! \int_0^{\infty}\!\! \frac{dy}{2}  &\langle  j_{x}(k, y)\varphi(k', y) +\varphi(k, y)j_{x}(k', y) \rangle, 
\end{align}
where $j_{x}(k, y) = - \sigma_{xy}(y) \partial_y \varphi(k, y)$, so that we again ignore the contribution of $\sigma_{xx}$ to the longitudinal current according to the assumptions made in the previous section.
Note that without a loss of generality we have written the energy flux for simplicity at $t = 0, x=0$. Rewriting the total potential $\vp(k, y)$ in terms of the amplitudes of charged and neutral modes and using $\langle \rho_i(k, \omega)\rho_j(k', -\omega)\rangle = 2\pi\delta_{ij}\delta(k - k') S_{ij}(k, \omega)$, we can rewrite the energy flux as 
\begin{equation}\label{flux}
J_E  = \frac{1}{2\sigma_{xy}} \int \frac{dk d\omega}{4\pi^2} \sum_{i=\{c,\sigma\}}v^2_iS_{ii}(k,\omega).
\end{equation}
In this case, the charged and neutral modes are orthogonal and the energy flux is a sum of the respective auto-correlation functions. However, generally and as it will become clear from the regime of $\nu = 2$, the heat current also has contributions from the cross-correlators between the modes as the eigenstates stop being orthogonal in the presence of dissipation. Nevertheless, all the correlations can be easily found within a unified approach based on the FDT. 
Before moving to that, we note that the above formula can be understood intuitively \cite{PhysRevB.85.075309}. The energy flux operator can also be viewed as a product of the energy density of the edge excitation and its velocity, which can be written as $\hat{J} = v\cdot \pi v \hat{\rho}^2(x,t)$. This indeed results in the expression \eqref{flux} upon taking a thermal averaging.

\section{Correlations from FDT\label{sec:FDT}}

A convenient method, allowing us to stay within a hydrodynamics formalism~\cite{PhysRevB.97.115418} relies on the FDT relation \cite{PhysRev.83.34, kubo}. Let us clarify how it can be applied in our system. First, we need to construct a perturbation Hamiltonian $\delta H(k, \omega)$ that would only excite\footnote{The perturbation can be very general, but we have already chosen to project our system on the two lowest modes and thus we concentrate only on them.} the low energy modes $\rho_{c, \sigma}(k)$.
For that we introduce the sources $\delta\vp_{1,2}(y)$ that couple to the total charge density
\begin{equation}
\delta H = -\int\! dy \Big[\rho_1(y) + \rho_2(y)\Big]\Big[\delta\vp_1(y) + \delta\vp_2(y)\Big].
\end{equation}
To use FDT and to compute the correlators for the charged and dissipative modes we need the perturbation to take the following form:
\begin{equation}
\delta H = - \rho_c \delta\vp_c - \rho_{\sigma} \delta\vp_{\sigma},
\end{equation}
where $\delta\vp_{c,\sigma}$ are the amplitudes of the sources that excite the corresponding modes. 
This form of the perturbation is achieved if we choose $\delta\vp_1(y) = \delta\vp_c$ and $\delta\vp_2(y) = \delta\vp_{\sigma} (2 y/\xi - 1)$. This particular choice has an interpretation. It means that the charged mode is excited by a spatially constant field, while the AG neutral mode -- by a linear field. The latter is consistent with our model, where we assume a constant electric field in the strip associated with this neutral mode. 
The FDT states that a correlator $S_{ij}(k,\omega)$ is simply related to the response function $\mc{G}_{ij} = \frac{\delta\rho_i}{\delta\vp_j}$ at a temperature $T \equiv 1/\beta$, where $\delta\rho_i$ describes a deviation of the charge density amplitude from its unperturbed value:
\begin{equation}\label{FDT}
S_{ij}(k,\omega) + S_{ji}(k,\omega) = \frac{2 \operatorname{Im} \{\mc{G}_{ij}(k,-\omega) + \mc{G}_{ji}(k,-\omega)\}}{1-e^{-\beta\omega}}.
\end{equation}
Meanwhile, the response functions $\mc{G}_{ij}$ can be obtained from the equations of motion \eqref{system} by introducing potentials $\delta\vp_{1,2}(y)$ into them. The result reads
\begin{equation}\label{loretzinans}
\mc{G}_{cc}\!=\! \frac{-k\sigma_{xy}}{k v_c\! -\! \omega\! -\! i\gamma}, \ \mc{G}_{\sigma\sigma} = \frac{-k\sigma_{xy} + 2i\sigma_{xx} \ve_0/v_{\sigma}}{k v_{\sigma}\! -\!\omega\! -\! 2i\sigma_{xx} \ve_0/\sigma_{xy}},
\end{equation}
where $\gamma$ is an infinitesimally small shift of the pole for the retarded response function. The contribution to the energy flux~\eqref{flux} from the charged mode is thus trivially described by the expected quantum of the heat flux
\begin{equation}
J_{cc} = \int_{-\infty}^{\infty} \frac{d\omega}{4\pi} \frac{\omega}{1-e^{-\beta\omega}} - J_{vac} = J_q,
\end{equation}
where we subtracted the vacuum part $J_{vac}\! = \!\int_0^{\infty} d \omega \omega/{4\pi}$. For the neutral mode on the other hand we obtain
\begin{equation}\label{Jnn}
J_{\sigma\sigma} = \int_0^T\frac{d\omega \omega}{2\pi} \int_{-1/\xi}^{1/\xi} \frac{dk\cdot v_{\sigma}  2 \sigma_{xx}\ve_0 /\sigma_{xy}}{(k v_{\sigma} + \omega)^2 + 4\sigma^2_{xx}\ve^2_0/\sigma^2_{xy}},
\end{equation}
where we introduced the ultra-violet cut-off for $k$ above which our theory breaks down. We also explicitly cut off the integral for $\omega$ by a small temperature controlled in the experimental setting.
To find $J_{\sigma\sigma}$ we note that the width of the Lorentzian defined by $\sigma_{xx} \ve_0 /\sigma_{xy}$ becomes larger than the region where the low energy physics is applicable. Thus, the integrand can be taken as a constant in the leading order in $\sigma_{xy}/\sigma_{xx}$ and the main contribution to the integral comes from $|k| \leq\pm 1/\xi$. Interestingly, all the auxiliary parameters such as the value of the cut-off $1/\xi$ are cancelled out and the energy flux carried by the AG mode is solely defined by the parameter $\sigma_{xy}/\sigma_{xx}$:
\begin{equation}\label{Jnn_result}
J_{\sigma\sigma} = \frac{\sigma_{xy}}{2\pi \sigma_{xx}} J_q
\end{equation}
This result is quite straightforward, since there is no coupling between the charged and AG neutral mode. In the next section we show that it is nevertheless robust to such coupling, namely at $\nu = 2$ where the two modes interact via an additional charge density. We remark once again that a breakdown of the heat current quantization is a consequence of the low-energy character of our model which imposes a cut-off $\pm 1/\xi$ in the integral \eqref{Jnn}. Formally, if the integral is extended to infinity, one restores the quantization.

\section{Energy flux  at $\nu = 2$\label{sec:2}}

\begin{figure}[h!]
\center
\includegraphics[scale = 0.5]{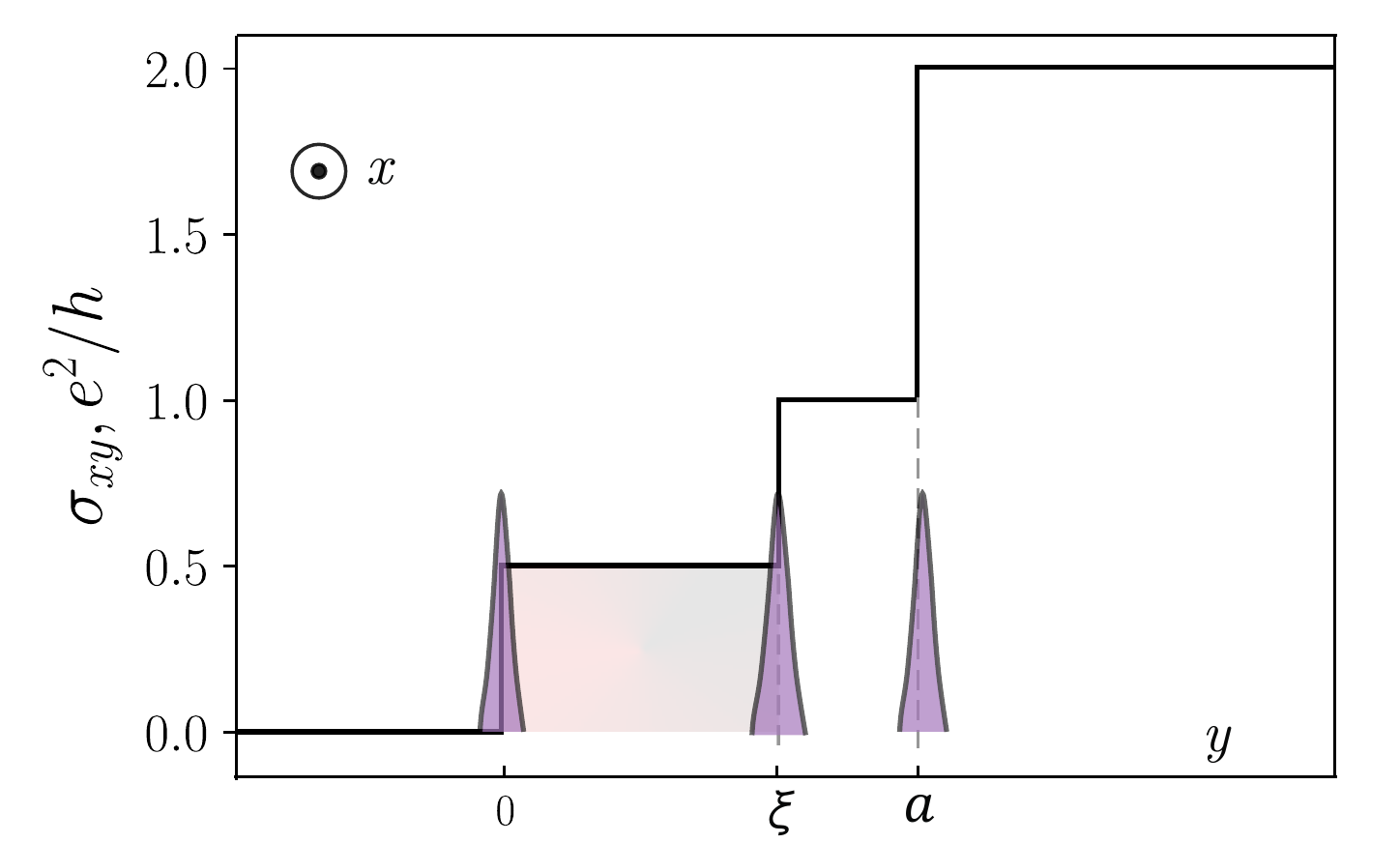}
\caption{A Hall conductivity profile of the edge cross-section at $\nu = 2$.
Again, as in Fig.~\ref{fig:density} the compressible strip is represented as a $1/2\cdot e^2/h$-step (shaded region). The charge densities accumulate at $y = 0; \xi; a$.
In this case all the eigenmodes are coupled due to the presence of dissipation and the geometry of the system. \label{fig:density2}}
\end{figure}

The system that we refer to in this section is depicted in Fig.~\ref{fig:density2}. We repeat the steps of the previous discussion
and first write down equations of motion. This time for three charge amplitudes $\rho_1(k,\omega), \rho_2(k,\omega), \rho_3(k,\omega)$ localized at $y = 0, y = \xi$ and $y = a$  respectively:
\begin{subequations}\label{EOM_nu2}
\begin{align}
&\omega \rho_1 = \frac{k \sigma_{xy}\vp(k,\delta)}{2} + i \sigma_{xx} \partial_y \vp(k, y)|_{\delta}, \\
&\omega \rho_2 = \frac{k \sigma_{xy}\vp(k,\xi)}{2} - i \sigma_{xx} \partial_y \vp(k, y)|_{\xi},\\
&\omega \rho_3 = k \sigma_{xy}\vp(k,a),
\end{align}
\end{subequations}
where the total potential is 
\begin{align}\label{phi_nu2}
\vp(k, y) = 2\pi \Big[ U(k|y|)\rho_1 + U(k|y-\xi|)\rho_2
 + U(k|y-a|)\rho_3\Big]
\end{align}
with the interaction kernel defined as before. We again use the condition of the constant electric field inside a strip, which results in $U'(y/D)|_{y = \delta}= U'( y/D)|_{y = \xi}$ and $U'(y/D)|_{y = a-\xi} = U'(y/D)|_{y = a}$. 
Diagonalizing then the system \eqref{EOM_nu2}, one finds that there are three non-orthogonal modes, two of which correspond to the usual charged and neutral modes, while the third one is an AG neutral mode residing in the strip. This mode again has a dominant dissipative contribution to its spectrum and we will be also referring to it as the {\it overdamped} one. The two other modes acquire a dissipative correction of $\mc{O}(k^2)$ due to the coupling with a dissipative mode and  for that reason we will be calling them {\it underdamped}. To show it analytically we first introduce the following velocities
\begin{align}
v_{1,2} &= 2\pi \sigma_{xy} \Big[U(\delta/D) \pm U(\xi/D)\Big]/2, \\
v_{3,4} &= 2\pi\sigma_{xy} \Big[U((a-\xi)/D) \pm U(a/D)\Big]/2.
\end{align}
and then expand the expressions for $v_{2,4}$ at small $\xi$ to the first order to get a relation
\begin{equation}
U'(y/D)|_{y = \delta}/ U'(y/D)|_{y = \xi} = v_2/v_4.
\end{equation}
It allows us to simplify the analytical treatment of the eigenvalues of the system \eqref{EOM_nu2}. Namely, the eigenmodes acquire the forms
\begin{align}
\omega_{c, n}(k) &=k\Big[ v_1+ \frac{v^2_2 - v^2_4}{2v_2} \pm \sqrt{v^2_3 + \frac{(v^2_2 - v^2_4)^2}{4 v^2_2}}\Big]\nonumber \\
 &- i k^2 \sigma_{xy}/\sigma_{xx}\cdot C_{c,n} , \\
\omega_{\sigma}(k) &= k(v_2 + v^2_4/v_2) - 2 i \frac{\sigma_{xx}}{\sigma_{xy}}\ve_0,
\end{align}
where $\ve_0 = 2 v_2/\xi$ as before.
The precise form of the positive constants $C_{c, n}$ is unimportant here since we never use the approximate expressions for the spectrum in the calculations below. But as an example, in the simplest case where $v_2 = v_4$, $C_{c,n} = v_2(v_1\pm v_3)/4\ve_0$. Exactly as mentioned above the charged and neutral modes acquire a dissipative correction, while for the overdamped mode the dissipative part remains unchanged compared to $\nu= 1$. It is not surprising, since the dissipative part represents the largest energy scale, so that the presence of low-energy modes is insignificant for it.
Additionally, the obtained spectrum of the neutral mode\footnote{In fact, when  $v_2 = v_4$, the ratio of the dissipative term to the linear one is the same both for the neutral and charged modes and equals to $ k \xi \sigma_{xy}/8\sigma_{xx}$.} is consistent with the experimental findings of Ref.~[\onlinecite{bocquillon}], where the quadratic behavior of the imaginary part of the wavevector was found at small $\omega$.

Intuitively, such spectrum suggests that the energy flux would have a similar structure compared to $\nu=1$: the AG mode would carry the same portion of the flux quantum, while the underdamped modes transfer together two flux quanta. Since the overdamped mode acts as a metal, to the leading order in $\sigma_{xy}/\sigma_{xx}$ it only renormalizes the velocities of the underdamped modes,  which do not enter the expression for the heat current. 

In agreement with the above arguments, we find indeed that  the  heat current carried  by all the modes is described by
\begin{equation}
J_E = \Big[2 + \frac{\sigma_{xy}}{2\pi\sigma_{xx}}\Big]  \pi T^2/12
\end{equation}
The details of the derivation can be found in the  appendix \ref{appA}. Here let us briefly describe the basic steps. The total expression for the heat flux has the form
\begin{align}\label{flux_nu2_def}
J_E = \frac{\sigma_{xy}}{2}\!\!  &\int\frac{dk d\omega}{(2\pi)^2}\Big[\frac{1}{2}\langle\varphi(k, 0)\varphi(k, 0)\rangle \nonumber\\
&+  \frac{1}{2}\langle\varphi(k, \xi)\varphi(k, \xi)\rangle + \langle\varphi(k, a)\varphi(k, a)\rangle\Big]
\end{align}
and is expressed in terms of all the auto- and cross-correlators $\langle\rho_i(k, \omega)\rho_j(k',\omega)\rangle$ via the relation \eqref{phi_nu2}. Upon finding these correlators through FDT, we then substitute them into the expression for the heat flux and calculate it to the leading order in $\sigma_{xy}/\sigma_{xx}$ by expanding our expressions for small $\omega$ based on $T\ll \sigma_{xx} \ve_0/\sigma_{xy}$. The lowest order term results in two flux quanta while the correction to it coincides with heat current $J_{\sigma\sigma}$ carried by the AG mode at $\nu = 1$, see Eq.~\eqref{Jnn_result}.

\section{Discussion\label{sec:disc}}

Motivated by the experiment [\onlinecite{pierre}], we have  proposed an effective model of dissipation at the edge of the integer quantum Hall system and found a corresponding heat current. In this model dissipation is associated with a compressible region arising in realistic experimental set-ups. Compressible regions were predicted to host an infinite number of hydrodynamic, or as we refer to AG modes\cite{AG}, which have never been observed. We argued that this might be due to the dissipation which have not been fully explored before. Building upon a low-energy theory and making a few assumptions such as the transport chirality and $\sigma_{xx}\gg \sigma_{xy}$ in the compressible strip, we have calculated the contribution to the heat current from the lowest AG mode at $\nu = 1$ and $\nu = 2$.

Starting with $\nu = 1$, we have discovered that the dominant contribution to the AG mode heat flux comes from all wavevectors $|k| \leq 1/\xi$. As a result this overdamped mode carries only a portion of the flux quantum, $(\sigma_{xy}/2\pi \sigma_{xx}) J_q$ -- consistent with the observation of Ref.~[\onlinecite{pierre}]. The answer is universal as it only contains the ratio $\sigma_{xx}/\sigma_{xy}$ while all extra parameters, such as velocities and high energy cut-off drop out. However, the exact numerical prefactor depends on the interaction details.

We generalize our result to $\nu = 2$, where the overdamped AG mode interacts with the charged and neutral modes. We find that the change in the low-energy spectrum of the AG mode is minimal and only its velocity changes. In addition, the charged and neutral modes acquire an imaginary $k^2$ contribution due to the interaction with the overdamped mode. We find that the heat current  they are transferring together is $(2 + \sigma_{xy}/2\pi\sigma_{xx}) J_q$. Moreover, the presence of the additional overdamped mode either at $\nu = 1$ or $\nu = 2$ results in the same correction to the heat current. Indeed, the contribution from the dissipative mode comes from the whole range of the wavevectors where the hydrodynamic description is justified. At this scale the low-energy modes can be considered as an electrostatic environment, whose presence can only lead to the renormalization of the velocity of the high-energy mode. As it does not enter into the heat current expression, the exact number of such slow modes does not matter.

Although the presented results are consistent with the experiments it is important to underline, that in the hydrodynamic limit, where AG mode exists, it is overdamped and carries less than a flux quantum. On the other hand a significant contribution to its heat current may come from the scales where our low-energy approach is not applicable. We thus conclude that in order to build up a complete picture of heat transport with dissipation one has to come up with a microscopic model, say of the impurities at the edge and go beyond the low-energy limit.

{\bf Acknowledgements.} We would like to thank Ivan Protopopov for the fruitful and instructive discussions. This work has been supported by the Swiss National Science Foundation.

\onecolumngrid
\appendix

\section{Heat current calculation at $\nu = 2$ \label{appA}}
Let us first rewrite the expression for the heat current \eqref{flux_nu2_def} in terms of the charge densities correlators:
\begin{align}\label{flux_from_dens}
J_E = \frac{1}{2\sigma_{xy}} \int \frac{dkd\omega}{2\pi} &\Big[(v_1^2 + v_3^2)S_{cc}+ \{v^2_3 + v^2_4 + (v_1 + v_2)^2\} S_{33} + (v^2_2 + v^2_4) S_{\sigma\sigma} \nonumber\\
& + v_3(2v_1 + v_2)(S_{c3} + S_{3c}) + v_3v_4 (S_{\sigma c}+S_{c\sigma}) + v_4(v_1 + 2v_2)(S_{\sigma 3} + S_{3 \sigma})\Big],
\end{align}
where $S_{ij} \equiv S_{ij}(k,\omega)$ and $ \langle \rho_i (k,\omega) \rho_j (k',-\omega)\rangle = 2\pi \delta(k - k')S_{ij}$. Here we use the following notations:
\begin{equation}
\rho_{c,\sigma} = \rho_2 \pm \rho_1,
\end{equation}
i.e. the $\rho_c$ and $\rho_{\sigma}$ denote the amplitudes of the charged and dissipative mode as they were for $\nu = 1$. In this way we see directly how the coupling with $\rho_3$ modifies the picture and can apply the same perturbation sources for the FDT in order to find the above correlators.
Indeed, the perturbation Hamiltonian can be written in the form
\begin{align}
-H_p &= \int dy \left(\frac{\rho_c}{2}[\delta(y) + \delta(y-\xi)] + \frac{\rho_{\sigma}}{2}[\delta(y-\xi)-\delta(y)] + \rho_3\delta(y-a)\right) \left(\delta\vp_c(y) + \delta\vp_{\sigma}(y) + \dvp_3(y)\right) \\
& = \frac{\rho_c}{2} \left(\dvp_1(0) + \dvp_1(\xi) + \dvp_2(0) + \dvp_2(\xi)\right) + \frac{\rho_{\sigma}}{2}\left(\dvp_1(\xi) - \dvp_1(0) + \dvp_2(\xi) - \dvp_2(0) + \rho_3 \dvp_3(a)\right).
\end{align}
Taking $\dvp_3(y)$ localized at $y = a$ with $\dvp_3(a)=\delta\vp_3$ and
$\dvp_1(y) = \vp_c, \ \dvp_2(y) = \vp_{\sigma} (2 y/\xi - 1)$ 
results in
\begin{equation}
H_p = -\rho_c\dvp_c - \rho_{\sigma} \dvp_{\sigma} - \rho_3 \dvp_3.
\end{equation}
We can now substitute these sources into the equations of motion \eqref{EOM_nu2} to find the linear responses. The obtained systems reads
\begin{equation}\label{EOM_sources}
\begin{pmatrix}
k v_1 - \omega & 0 & k v_3 \\ 
0 & k v_2 - \omega - 2 i\frac{\sigma_{xx}}{\sigma_{xy}} \ve_0 & k v_4  - 2 i\frac{\sigma_{xx}}{\sigma_{xy}} \frac{ v_4}{v_2} \ve_0 \\ 
k v_3 & k v_4 & k (v_1 + v_2) - \omega
\end{pmatrix}\begin{pmatrix}
\delta\rho_c \\ 
\delta\rho_{\sigma} \\ 
\delta\rho_3
\end{pmatrix} =
\begin{pmatrix}
-k\sigma_{xy} \dvp_c \\ 
-(k\sigma_{xy}  - 2 i\frac{ \sigma_{xx}}{v_2} \ve_0)\dvp_{\sigma} \\ 
-k \sigma_{xy}\dvp_3
\end{pmatrix}.
\end{equation}
The nine linear responses $\delta\rho_i / \delta\rho_j$ can now be found straightforwardly:
\begin{align}\label{linear_responses}
&\frac{\delta\rho_c}{\delta\vp_c}  = - \frac{k\sy \Big[v_2(k v_2 - \omega - 2i\sx /\sigma_{xy}\ve_0)(k(v_1+v_2)-\omega)-k v^2_4(k v_2-2 i \sx/\sigma_{xy}\ve_0)\Big]}{v_2\Delta}, \\
&\frac{\delta\rho_c}{\dvp_3} = \frac{k^2 v_3\sigma_{xy}(k v_2 - \omega -2i\sigma_{xx}/\sigma_{xy}\ve_0)}{\Delta} = \frac{\delta\rho_3}{\delta\vp_c}, \quad 
\frac{\delta\rho_c}{\dvp_{\sigma}} = -\frac{k^2v_3 v_4\sigma_{xy}(kv_2 - 2i\sigma_{xx}/\sigma_{xy}\ve_0)}{v_2\Delta}\\ 
&\frac{\delta\rho_{\sigma}}{\delta\vp_c} =  -\frac{k^2v_3v_4\sigma_{xy}(kv_2 - 2i\sigma_{xx}/\sigma_{xy}\ve_0)}{v_2\Delta},\\
&\frac{\delta\rho_{\sigma}}{\delta\vp_{\sigma}} = -\frac{\sigma_{xy}(k v_2 - 2i\sigma_{xx}/\sigma_{xy}\ve_0)}{v_2\Delta} \Big[(k v_1 - \omega)(k(v_1 + v_2)-\omega)-k^2v^2_3\Big], \\
&\frac{\delta\rho_{\sigma}}{\delta\vp_3} = \frac{k v_4\sigma_{xy} (k v_1 - \omega)(k v_2 - 2i\sigma_{xx}/\sigma_{xy}\ve_0)}{v_2\Delta}, \\
&\frac{\delta\rho_3}{\delta\vp_{\sigma}} = \frac{k \sigma_{xy}v_4 (k v_1 - \omega)(k v_2 - 2i\sigma_{xx}/\sigma_{xy}\ve_0)}{v_2\Delta}, \\
 &\frac{\delta\rho_3}{\delta\vp_3} = -\frac{k\sigma_{xy}(k v_1 - \omega)(k v_2 - \omega-2i\sigma_{xx}/\sigma_{xy}\ve_0)}{\Delta},
\end{align}
where the determinant $\Delta(k,\omega)$ is described by
\begin{align}\label{Delta}
\Delta(\omega, k) = & k^3 \{v^2_1 v_2 - v_2 v^2_3 + v_1 (v^2_2 - v^2_4)\} + 
 k^2 (-v^2_1 - 3 v_1 v_2 - v^2_2 + v^2_3 + v^2_4) \omega + 
 2 k (v_1 + v_2) \omega^2 - \omega^3 +\\
 & i \sigma_{xx} /\sigma_{xy} \ve_0( - 2 k^2 v_1 v_2  + 
   2  k^2 (v^2_3 -v^2_1) + 2  k^2 v_1 v^2_4/v_2  + 
   4  k v_1  \omega + 2  k v_2  \omega - 
   2  k v^2_4/v_2  \omega - 2 \omega^2).
\end{align}
Again using the FDT relation \eqref{FDT}, we find the correlation functions and substitute them in the integral~\eqref{flux_from_dens}. 
Although the acquired relations are cumbersome the structure of the integral is clear. There are three poles corresponding to the three modes and the contribution from the low-energy charged and neutral excitations can be separated from the dissipative mode. We thus perform the calculations in the following manner. First, we rewrite 
$\frac{1}{\Delta(k,\omega)} = \Big[\operatorname{Re} \Delta - i \operatorname{Im} \Delta\Big]/\Big[(\operatorname{Re} \Delta)^2 + (\operatorname{Im} \Delta)^2\Big]$ and note that the imaginary part $\operatorname{Im} \Delta \propto \sigma_{xx}$ and therefore dominates over $\operatorname{Re} \Delta$ for low enough $k$. It allows us to simplify this expression further, ending up with
\begin{equation}\label{simplified_delta}
\frac{1}{\Delta(k,\omega)} \approx - \frac{i}{\operatorname{Im} \Delta}+\frac{\operatorname{Re} \Delta}{(\operatorname{Im} \Delta)^2} .
\end{equation}
The first term here is proportional to $1/\sigma_{xx}$ and it is this one that results in the two flux quanta. Let us  clarify it. 

First note, that $\operatorname{Im} \Delta = \mc{O}(k^2)$ and contains two poles corresponding to the charged and neutral modes. Therefore, the imaginary term in \eqref{simplified_delta} multiplied by the leading order terms of $k$ and $\omega$ in the numerators of linear response functions \eqref{linear_responses} brings us to the two flux quanta. One can easily understand that it is the case, since all the leading order terms in the aforementioned numerators are also proportional to $\sigma_{xx}$, which hence drops out. One is left then with a simple integration over $k$ of two delta functions arising from the two poles. 

To obtain a correction to this result one needs to multiply $- \frac{i}{\operatorname{Im} \Delta}$ by the next order terms of $k$ and $\omega$ in the numerators and adding to that the product between $\frac{\operatorname{Re} \Delta}{(\operatorname{Im} \Delta)^2}$ and the lowest order terms in $k$ and $\omega$. In the end, one makes an expansion in $\omega$, since it is cut off by small $T$, while the main contribution in $k$ comes from the ultra-violet cut-off $1/\xi$ as we saw in Sec.~\ref{sec:FDT}.
Following this strategy one can discover the correction $\sigma_{xy}/2\pi\sigma_{xx} J_q$ -- same as at  $\nu = 1$.


\twocolumngrid

\bibliography{draft_v1}
\bibliographystyle{apsrev4-1}
\end{document}